\documentclass[preprint,showpacs,preprintnumbers,amsmath,amssymb,nofootinbib]{revtex4}

\usepackage{amsmath}
\usepackage{amsfonts}
\usepackage{amssymb}
\usepackage{axodraw}

\newcommand{\GeV}{{\rm GeV}}

\newcommand{\bear}{\begin{eqnarray}}	
\newcommand{\eear}{\end{eqnarray}}	
\newcommand{\beqstar}{\begin{eqnarray*}}	
\newcommand{\eeqstar}{\end{eqnarray*}}

\begin{document}

\preprint{HUTP-03/A009}

\title{Pseudonatural Inflation}

\author{Nima Arkani-Hamed, Hsin-Chia Cheng, Paolo Creminelli and Lisa Randall}
\affiliation{Jefferson Physical Laboratory, \\
Harvard University, Cambridge, MA 02138, USA\\}

\begin{abstract}
We study how to obtain a sufficiently flat inflaton potential that is natural from the particle physics
point of view. Supersymmetry, which is broken during inflation, cannot protect the potential against 
non-renormalizable operators violating slow-roll. We are therefore led to consider models based on non-linearly 
realized symmetries. The basic scenario with a single four-dimensional pseudo 
Nambu Goldstone boson requires the spontaneous breaking scale to be above the Planck scale, 
which is beyond the range of validity of the field theory description, so that quantum gravity corrections are not 
under control. A nice way to obtain consistent models with large field values  is to consider simple extensions 
in extra-dimensional setups. 
We also consider the minimal structures necessary to obtain purely four-dimensional models with spontaneous 
breaking scale below $M_P$; we show that they require an approximate symmetry that is supplemented by 
either the little-Higgs mechanism or supersymmetry to give trustworthy scenarios.  
\end{abstract}


\maketitle

\section{\label{sec1}Introduction}
Inflation is surely the most compelling paradigm for solving many problems of the standard big bang 
cosmology~\cite{Guth:1980zm,Linde:1981mu,Albrecht:1982wi}. 
Besides its theoretical appeal, its basic predictions of a flat Universe with a nearly scale-invariant
spectrum of adiabatic perturbations are now experimentally well tested by the Cosmic Microwave Background 
Radiation (CMBR) anisotropies and the Large Scale Structures (LSS) galaxy surveys. 

The basic framework can be realized in models as simple as a single scalar field with a monomial potential.
Although such simple toy-models can be attractive, they are tremendously unnatural from the particle physics
point of view. In scenarios where the inflaton takes values above the Planck mass 
($M_P = (8\pi G)^{-1/2}$)~\cite{Linde:gd}, 
the use of a simple potential requires the fine-tuning of an infinite number of non-renormalizable operators, 
suppressed by powers of $M_P$. 

The inflaton potential must be sufficiently flat to allow a slow-rolling phase, but at the same time it must 
couple to other fields to provide an efficient reheating and, in hybrid models, to trigger the final phase 
transition. 
There are only two known candidates for keeping a scalar potential nearly flat and stable under radiative 
corrections: supersymmetry (SUSY) and non-linearly realized symmetries. The latter mechanism applies
both to a pseudo Nambu-Goldstone boson (PNGB) and to the extra components of gauge fields propagating
in extra dimensions; both are protected at lowest order by a shift symmetry.

So far most of the attention in inflation model-building has been devoted to supersymmetry, but this symmetry
alone cannot naturally provide potentials that are flat enough for inflation, once supergravity effects
are included. In the next section (section \ref{sec:sol}), we will describe this problem and review the proposed 
solutions. None of them is completely compelling.

In section~\ref{sec:idea} we turn our attention to the other, much less studied, candidate: the shift symmetry. 
Even if the Goldstone theorem ready provides flat directions for inflation, it is not trivial to build 
inflationary models based on PNGBs, essentially because both the potential and its slope vanish in the limit 
in which the explicit breaking is turned off. 
The simplest scenario with a single PNGB does not work unless the symmetry breaking scale is higher than the 
Planck scale, which is presumably outside the range of validity of an effective field theory description. Moreover
it is expected that quantum gravity effects will explicitly break the global symmetry, giving a typical
scale for the potential of order $M_P$, far too big to satisfy the COBE constraint. 

We show that these problems are not present in theories with extra dimensions. In particular,
the extra components of gauge fields living in extra dimensions provide natural candidates 
for the inflaton \cite{Arkani-Hamed:2003wu}.

In the rest of the paper we concentrate on the requirements for building purely 4d models with PNGBs, 
with a symmetry breaking scale below $M_P$.
This requires more complicated structures such as hybrid inflation models~\cite{Linde:1993cn}. 
In section~\ref{sec:models} we discuss the necessary ingredients for building natural 4d models. The PNGB 
potential of the inflaton needs protection from the interactions which are required to end
inflation and to reheat the Universe. We present a SUSY model as well as non-SUSY models based on the same 
recent ideas which were used to build new models of electroweak symmetry breaking. Some of the details are 
left to the Appendix.
We draw our conclusions in section \ref{sec:conclusions}.

\section{\label{sec:sol} Higher dimension operators and the SUGRA $\eta$ problem}
As noted in the introduction, non-renormalizable operators are clearly very crucial in 
models of inflation in which the inflaton variation is bigger than the Planck scale, because
they are naively more important than lower dimension operators. This makes it very hard to 
justify any 4d model with a big variation of field values. We will discuss in the next section how 
this problem can be solved in models in which the 4d effective field theory is the dimensional
reduction of an higher dimensional theory.

Non-renormalizable operators are important also in theories where the inflaton variation is much
smaller than $M_P$. This is clear if one considers operators of 
dimension 6, which can give a mass term 
\begin{equation}
\label{eq:correct}
\frac{V}{M_P^2} \phi^2 \sim H^2 \phi^2
\end{equation}
to the inflaton, spoiling slow-roll. 

One would think that supersymmetry can provide flat directions for inflation in a rather natural way; 
however, it is known that this is not quite true once supergravity corrections are included 
\cite{Copeland:1994vg}. Non minimal terms in the K\"ahler potential can obviously 
give contributions like (\ref{eq:correct}), but the same kind of corrections are present also with 
a minimal K\"ahler potential. 
The supergravity potential, neglecting for the moment the D-term
contribution, can be expressed as a function of the K\"ahler potential
$K$ and the holomorphic  superpotential $W$ as 
\begin{equation}
\label{eq:SUGRApot}
V = e^{K/M_P^2} \left[(K^{-1})^i_j L_i L^j - 3
\frac{|W|^2}{M_P^2}\right] \;, 
\end{equation} 
where $L_i \equiv W_i + K_i \cdot \frac{W}{M_P^2}$. During inflation supersymmetry 
is broken because the vacuum energy is positive. Taking, at the lowest order, a 
canonically normalized K\"ahler potential  $K = \phi^* \phi$, the exponential
factor in front of $V$ gives a  mass to any flat direction of order
$V/M_P^2 \sim H^2$. This point is quite clear in the superconformal formalism, where
the kinetic term for $\phi$ can be expressed using a superconformal compensator $\Phi$
as 
\begin{equation}
\label{eq:superconformal}
\int \!d^2 \theta d^2\bar\theta \;\; \Phi \Phi^\dagger \phi \phi^\dagger \;.
\end{equation}
As $\Phi \Phi^\dagger$ contains the Ricci scalar, we obtain a non-minimal coupling of $\phi$ 
to gravity which gives the mass correction during inflation.

This effect gives a tilt to the inflaton potential and it
is simple to check that its contribution to the slow-roll $\eta$
parameter ($\eta \equiv M_P^2 V''/V$) is exactly 1, while a slow-roll
phase requires $\eta \ll 1$. There can be additional contributions in the
potential (\ref{eq:SUGRApot}) which are of the same order of
magnitude and a cancellation is possible. Nevertheless this required 
cancellation introduces a fine-tuning problem, which is often referred to as the
$\eta$-problem. The on-going experiments on the CMBR and on LSS are
making the problem increasingly acute. A conservative limit on the
spectral index is now $|n-1|< 0.1$, which turns into a limit for
$\eta$: $\eta < 0.05$.  Unless a better reason for the cancellation is
found, a fine-tuning of at least $1/20$ is required.  The situation
somewhat resembles the Higgs hierarchy problem: the top Yukawa and the
gauge and quartic  couplings would drive the Higgs mass towards the
scale $\Lambda$ where new physics comes in, but a  certain separation
is required to account for electroweak precision tests. Here gravity
itself drives the inflaton mass towards $H$, but again the two scales
must be separated to allow a sufficient  amount of inflation.

Several ways to overcome the problem have been proposed. Before reviewing
them, we want to stress that none of them is entirely convincing. Most of
them rely on assumptions about the fundamental theory, which
cannot be justified from the effective low energy point of view. This
is not better than assuming a certain  cancellation among the
various terms in (\ref{eq:SUGRApot}). Below we discuss some proposed 
solutions to the $\eta$-problem.
For additional references see \cite{Lyth:1998xn}.

{\bf Superpotential linear in the inflaton} \cite{Copeland:1994vg}. It
is easy to verify that in this case  the contribution to $\eta$ coming
from the exponential factor in (\ref{eq:SUGRApot}) is
canceled. However, one is left to assume a small quartic term $(\phi
\phi^*)^2$ in $K$. Even if the situation might be considered better than 
in the general case, the fine-tuning problem is still there, as there is no 
symmetry which can protect the smallness of this term. 

{\bf Particular form of the K\"ahler potential}. If $T$ is a modulus
(e.g., a compactification radius) and $\phi$  is the inflaton and the K\"ahler
potential depends only on the combination  $\rho \equiv (T +T^* - \phi^*
\phi)$ (e.g., $K = -3 \log \rho$), a flat direction is preserved
\cite{Gaillard:1995az}. This could be ensured by a so-called
Heisenberg symmetry and it seems to be quite generic  in orbifold
compactifications of superstrings. The problem is that, during
inflation, one also gets a runaway  potential in the $\rho$
direction. It is hard to justify why a stabilization mechanism should
depend on  the $\rho$ variable and not on $T$ itself, as the Heisenberg
symmetry is not a symmetry of the full theory. On the other hand, if $T$ is
stabilized, corrections to the inflaton potential are  reintroduced,
giving $\eta \sim 1$.  All this kind of solutions relies on particular
features of the K\"ahler potential, which cannot be justified in term
of symmetries of the low energy theory, but must be taken from the UV
stringy  completion.

{\bf D-term inflation}. In addition to the F-term potential
(\ref{eq:SUGRApot}), D-term contributions are also present: 
\begin{equation}
\label{eq:Dterm}
V_D = \frac{g^2}2 {\rm Re} f_{AB}^{-1} D^A D^B \;,\qquad D^A = K^i
(T^A)_i^j \phi_j + \xi^A \;, 
\end{equation}
where $f$ is the gauge kinetic
function, $T^A$'s are the gauge group generators and $\xi$ is a
Fayet-Iliopoulos  (FI) term, which is admissible only for U(1)
groups. If the vacuum energy during inflation is dominated  by a
D-term, the $\eta$-problem is simply not there
\cite{Binetruy:1996xj}. One can easily build a hybrid  inflation model
taking the inflaton to be a neutral superfield $S$, coupled to two
charged multiplets  $\phi_+$ and $\phi_-$ through a superpotential $W
= \lambda S \phi_+ \phi_-$. For large values of  the scalar component
of $S$, $\phi_+$ and $\phi_-$ are stuck at the origin, so that the
gauge  symmetry is unbroken and the vacuum energy is dominated by the
FI term. For smaller values, the  negatively charged scalar becomes
tachyonic and we go to a vacuum where the U(1) is broken and the
vacuum energy is zero. The $S$ direction is classically flat, but it
is lifted by quantum corrections as  supersymmetry is broken. 
The potential is generically
flat enough  to allow slow-roll and no
$\eta$-problem seems to be present. However, after a closer look,
it seems difficult to  get a viable
scenario of inflation both with an anomalous U(1)
and with a non-anomalous one.

\begin{enumerate}
\item Anomalous U(1) with Green-Schwarz mechanism of anomaly
cancellation \cite{Dine:xk}.  As in this case the non-linear
transformation of the dilaton cancels the anomaly, its behavior is
clearly crucial: during inflation the dilaton gets a runaway potential
and it must be stabilized.  The stabilization mechanism generically
gives F-term contributions bigger than D-terms
\cite{Arkani-Hamed:1998nu}, thus reintroducing the $\eta$-problem.



\item An explicit FI term for a non-anomalous U(1) is
introduced. Supergravity requires this U(1) to be an  R-symmetry
\cite{Barbieri:1982ac}. This
point is often overlooked in the literature about inflation.\footnote{
We thank M. Luty for stressing this point to us.}
In the superconformal formalism it is easy
to understand why the naive extension of the rigid FI term is not
gauge invariant by itself. With a compensator chiral multiplet $\Phi$,
a rigid FI  term would be promoted to $[\Phi \bar \Phi \xi V]_D$, which
is no longer invariant under $V \rightarrow V + \Lambda + \bar
\Lambda$ since $[\Phi \bar \Phi \Lambda]_D$ and $[\Phi \bar \Phi \bar
\Lambda]_D$ are non-zero. To write a gauge-invariant generalization of
a FI term, $\Phi$ must transform under the $U(1)$ symmetry: a term of
the form $[\Phi \bar \Phi e^{\xi V}]_D$ will be invariant, provided the
compensator undergoes a super-Weyl transformation $\Phi \rightarrow
e^{-\xi \Lambda} \Phi$, $\bar \Phi \rightarrow e^{-\xi \bar \Lambda}
\bar \Phi$ \cite{Barbieri:1982ac}.  This implies that the U(1) symmetry
must in fact be a gauged version of the U(1)$_R$ symmetry: the
gravitino must be charged.

The framework is now quite constrained: taking the normalization in
which the charge  of the gravitino is $-1$, the vacuum energy during
inflation is given by 
\begin{equation}
\label{eq:Dvacuum}
V_0 = \frac{g^2}2 \xi^2 = \frac{g^2}2 (2 M_P^2)^2 \;;  
\end{equation}
where the FI term is fixed by the gravitino charge.
The R-symmetry requires that the superpotential has charge $+2$: as the
scalar component of $S$  must be neutral, $\phi_+$ and $\phi_-$ have
charge $2+Q$ and $-Q$ respectively. The classically flat direction $S$
is lifted by quantum corrections and it gets a potential\footnote{Note
that the effective potential for $S$ is different from the case
\cite{Binetruy:1996xj} of a U(1) which  is not an R-symmetry, because
here the charges of $\phi_+$ and $\phi_-$ do not add up to zero and
an additional term $\propto |S|^2 \log (|S|^2/\mu^2)$ is
induced. Anyway its contribution is at most  comparable to the $\log
|S|^2$ piece, so that neglecting it does not alter the conclusions.}:
\begin{equation}
\label{eq:Dlifting}
V_{\rm 1-loop} = \frac{g^2 \xi^2}{2} \left(1 + \frac{2+2Q+Q^2}{8
\pi^2} g^2 \log\frac{\lambda^2  |S|^2}{\mu^2}\right) \;, 
\end{equation} 
where
$\mu$ is the renormalization scale. It is easy to obtain the
constraint from the COBE normalization (see e.g. \cite{Lyth:1998xn}),
which is independent of the coupling constant $g$ and requires a huge
charge $Q$: 
\begin{equation}
\label{eq:Dnorm}
\sqrt{\frac{\xi}{Q}} \simeq 10^{16} \GeV \quad \Rightarrow \quad Q
\sim 10^6 \;.  
\end{equation} 
This is technically natural but quite unreasonable, 
especially because anomaly cancellation gives 
strong contraints on the spectrum \cite{Chamseddine:1995gb,Castano:1995ci}.  
Even if we allow additional contributions to lift the inflaton potential, the  vacuum
energy (\ref{eq:Dvacuum}) is clearly too big to satisfy the COBE
normalization unless $g$ is very small.
\end{enumerate}

In summary, due to the large vacuum energy during inflation, supersymmetry
is badly broken in such a way that it lost the power to preserve the
flat direction required by inflation. We need some other mechanism
to obtain a natural flat potential for inflation.

\section{\label{sec:idea}shift symmetries I: general discussion}

If we want to explain the lightness of the inflaton
from the low energy point of view, we must rely on symmetry
arguments. Supersymmetry alone is insufficient, as we explained above,
so that one is naturally led to consider approximate bosonic shift
symmetries: i.e., the inflaton as a pseudo  Nambu-Goldstone Boson.
This is certainly not a new idea. Models of inflation based on
PNGBs were discussed in \cite{Freese:1990rb,Adams:1992bn} and in many subsequent
works\footnote{An approximate symmetry has also been used to have light fields, different 
from the inflaton itself, during inflation. This is important for the curvaton scenario 
\cite{Lyth:2001nq,Moroi:2001ct} or for moduli fields~\cite{Adams:1996yd}.}. 
In this paper we want to emphasize that this seems to be the only natural way to keep 
the potential flat for a slow-roll inflation. 

However, it is not straightforward to obtain satisfactory models. 
The original ``natural inflation'' model is based on a single PNGB field 
parametrized by an angular variable $\theta \sim \theta + 2\pi$. In the limit of exact symmetry
$\theta$ is a flat direction. With the addition of an explicit breaking term the 
Lagrangian is of the form
\begin{equation}
\label{eq:PNGBlag}
{\cal{L}} = \frac{1}{2} f^2 (\partial \theta)^2 - V_0 (1 - \cos\theta) \;, 
\end{equation} 
where $f$ is the spontaneous breaking scale. The canonically normalized
field is $\phi = f \theta$, so the potential is naturally a function of
$\phi/f$, which can be flat for large $f$.
This scenario is however problematic, because the requirements $\epsilon \ll 1$ and $\eta \ll 1$ 
on the slow-roll parameters
\begin{equation}
\label{eq:slow}
\epsilon \equiv \frac{M_P^2}{2}\left(\frac{V'}{V}\right)^2 \sim
\frac{M_P^2}{f^2} \;, \qquad \eta \equiv M_P^2 \frac{V''}{V} \sim
\frac{M_P^2}{f^2}
\end{equation}
gives $f \gg M_P$. 

If we interpret $f$ as some symmetry breaking vacuum expectation value
(VEV), then this would require that the field theory is valid above $M_P$, which is hard to justify.
But the real problem is that we expect that quantum gravity effects, such as the virtual appearance of black holes,
will explicitly break the approximate symmetry\footnote{Quantum gravity effects on a PNGB potential are 
known to be dangerous in the case of the axion. See e.g.~\cite{Kallosh:1995hi}.}. These effects, usually 
suppressed by powers of $f/M_P$, are here unsuppressed\footnote{Naive dimensional analysis 
suggests, in the limit of strong coupling gravity, that higher dimension operators arising from quantum gravity 
effects are suppressed by $M_P = (8 \pi G)^{-1/2}$ and not by the alternative definition of the Planck mass 
$G^{-1/2}$.}, so that it is hard to justify why $V_0$ should be smaller than $M_P$, as required by the COBE 
bound on the overall normalization of density perturbations: $\delta \rho/\rho \sim 10^{-5}$. It is the same 
problem of higher dimension operators we discussed in the previous section: here the inflaton variation is 
bigger than $M_P$ so that non-renormalizable operators are important. Quantum gravity effects will induce 
higher-dimension operators which badly break the symmetry, changing the potential in (\ref{eq:PNGBlag}).  
Therefore a single PNGB in a 4d field theory with the simple potential in (\ref{eq:PNGBlag}) 
cannot provide a satisfactory model of inflation.

The situation is changed when we consider theories with extra dimensions \cite{Arkani-Hamed:2003wu}. 
If the 4d effective field theory description comes from the dimensional reduction of a higher dimensional 
theory, it is possible to build models with variation of the inflaton field bigger than $M_P$, while
keeping the effects of higher-dimension operators under control.  
Locality in extra dimensions can in fact prevent large corrections to the inflaton potential from 
quantum gravity effects. 

Consider a 5d model with the extra dimension compactified on a circle of radius $R$. The extra component
$A_5$ of an abelian gauge field propagating in the bulk cannot have a local potential, due to the higher
dimensional gauge invariance; a shift symmetry protects it similarly to what happens to a four-dimensional 
PNGB.
A non-local potential as a function of the gauge invariant Wilson loop
\begin{equation}
\label{eq:WL}
e^{i \theta} = e^{i \oint A_5 d x^5}
\end{equation}
will however be generated in presence of charged fields in the bulk. At
energies below $1/R$, $\theta$ is a 4d field with a Lagrangian of the form
\begin{equation}
{\cal L} = \frac{1}{2\,g_4^2 (2 \pi R)^2} (\partial \theta)^2 - V(\theta) +
\cdots
\end{equation}
where $g_4^2 = g_5^2/(2 \pi R)$ is the 4D gauge coupling, and the potential
$V(\theta)$ is given at one-loop by
\cite{Hosotani:1983xw,Hatanaka:1998yp,Antoniadis:2001cv,vonGersdorff:2002as,Cheng:2002iz}
\begin{equation}
\label{eq:WLpot}
V(\theta) = - \frac{1}{R^4} \sum_I (-1)^{F_I} \frac{3}{64 \pi^6}
\sum_{n=1}^{\infty}
\frac{\cos(n q \theta)}{n^5} \;,
\end{equation}
where $F_I=0 (1)$ for massless bosonic (fermionic) fields of charge $q$
coupled to $A_5$. Note that the potential is of the same form as in natural inflation 
(with small corrections from additional terms in the sum), with the effective decay 
constant given by
\begin{equation}
\label{eq:decay}
f_{\rm eff} = \frac{1}{2\pi g_{\rm 4d} R} \;.
\end{equation}
It is easily seen that $f_{\rm eff}$ can be bigger than $M_P$ for sufficiently
small $g_{\rm 4d}$; the slow-roll condition $f_{\rm eff} \gg M_P$
requires only that
\begin{equation}
\label{eq:ourslow}
2 \pi g_{\rm 4d} M_P R \ll 1 \;.
\end{equation}
The canonically normalized field is $ \phi = \theta f_{\rm eff}$. Due to the higher
dimensional nature of the model, the potential (\ref{eq:WLpot}) can be trusted even
when the 4d field $\phi$ takes values above $M_P$; no dangerous
higher-dimension operator can be generated in a local higher-dimensional theory. 
This conclusion is quite important as it is commonly believed
that any inflation model with field values above $M_P$ cannot be justified from a particle
physics point of view; we see that this conclusion is valid only if we restrict to purely
4d models. Quantum gravity corrections to the potential (\ref{eq:WLpot}) are negligible if the extra
dimension is bigger than the Planck length, different from what is expected in a 4d PNGB model. 
Again locality in the extra space is the key feature; virtual black
holes cannot spoil the gauge invariance and do not introduce a local potential
for $A_5$, while non-local effects are exponentially suppressed by $\sim e^{- 2 \pi M_5 R}$, because 
the typical length scale of quantum gravity effects (the 5d Planck length $M_5^{-1}$) is much 
smaller than the size of the extra dimension.

It is worthwhile stressing that a variation of the inflaton field during inflation bigger than
$M_P$ is required to have a significant and measurable production of gravitational waves \cite{Lyth:1996im}. 
It seems that the only way to get a realistic scenario of this kind is in an extra-dimensional setup.

Another example using extra dimensions is the idea of ``brane inflation''
\cite{Dvali:1998pa}.  Also this model can be considered based on a PNGB. 
In fact, the inflaton is the field which describes the distance between two branes. 
It is massless in the limit in which we neglect the interactions between the two branes, 
because it is the Goldstone boson of the broken translational invariance. 
The non-trivial potential generated by the interactions between the two branes has to be 
very flat when two branes are far apart, again by the locality in extra dimensions. From
the 4d point of view the inflaton takes values above the Planck scale, but the extra dimensional
completion allows to control higher-dimension operators. Moreover quantum gravity
effects are again suppressed by locality, which is really the key ingredient of this type of models.

One could ask whether it is possible to derive  a purely
4d theory from the simple 5d model based on the Wilson line 
by applying the recent idea of deconstructing dimensions, 
where the Wilson line in the extra
dimension corresponds to a 4d PNGB~\cite{Arkani-Hamed:2001ca,Hill:2000mu,Cheng:2001vd,Arkani-Hamed:2001nc}.
In this case one replaces the 5d gauge theory by a chain of 4d gauge groups,
with the adjacent gauge groups connected by the link fields, which get
nonzero VEVs and break the gauge groups down to the diagonal one.
There is one linear combination of the Nambu-Goldstone bosons not eaten by the massive gauge fields.
It remains light and corresponds to the non-local Wilson line field in the 5d case.
However the symmetry breaking scale,
$f_{\rm link} = \sqrt{N} f_{\rm eff}$,
where $f_{\rm link}$ is the VEV of the link fields and $N$ is the number of
the sites, is still required to be bigger than $M_P$. 

In the rest of the paper we concentrate on 4d models, 
by which we mean that there is no (gravitational) extra dimension with 
size larger than the Planck length and the theory is 4-dimensional all the
way up to the 4d $M_P$.
In this case we can not use the locality in extra dimensions to protect
the flat inflaton potential and it is only sensible to consider
$f \ll M_P$. 
As explained, this is not consistent with slow-roll in a scenario with the simplest PNGB 
potential, so that one is naturally led to consider models which involve more than one field. 
With $f \ll M_P$ the corrections to the inflaton potential due to quantum gravity effects
can be sufficiently suppressed if the explicit symmetry breaking operators arising from
quantum gravity are prohibited up to a high dimension. Operators of dimension six contributing
directly to the inflaton mass are still dangerous because their effect on the mass can be 
of order $V/M_P^2 \sim H^2$. 
We would like to emphasize that quantum gravity corrections crucially depend on the UV completion of 
each model {\em below} the Planck scale. There are many ways to suppress quantum gravity effects. 
Besides locality in the extra space discussed above, additional (discrete or continuous) gauge   
symmetries in the UV theory can forbid dangerous operators. 
For example, in a dimensionally deconstructed gauge theory with many sites, the PNGB is the product 
of many link variables, so that the only gauge invariant operators are of very high dimensions and
the Planck scale effects are suppressed by many powers of $f/M_P$~\cite{Hill:2002kq}.

\section{\label{sec:models}Shift Symmetries II: Field Values smaller than $M_P$}

To obtain trustworthy 4d inflation models we must require
the symmetry breaking scale $f$ and field values
smaller than $M_P$ so that the simple potential (\ref{eq:PNGBlag}) does not work.
A more complicated PNGB potential is needed. In particular, the variation of the
potential during inflation and the total height of the potential should
be controlled by different terms with different scales. A sharp drop of
the potential is therefore needed at some point to end inflation. However,
such sharp drop in the potential explicitly breaks the shift symmetry of
the inflaton field and may spoil the flatness of the inflaton potential
through radiative corrections. We need to examine this point in more detail.

Let us first consider the case of single field inflation. Being a PNGB, its
potential is periodic. To have a separation of scales inflation must occur near the maximum 
so that the potential is sufficiently flat, otherwise we come back to the requirement of natural inflation 
$f \gg M_P$.
Near the maximum (chosen to be at $\psi=0$) we can expand the potential 
\begin{equation}
\label{eq:maximum}
V(\psi) = V_0 -\frac{m^2}{2} \psi^2 -\frac{\lambda}{4} \psi^4 + \cdots \;.
\end{equation} 
If $V_0 \sim m^2 f^2$, where $f$ is the symmetry
breaking scale and is also the maximal variation of $\psi$, as in the case 
of (\ref{eq:PNGBlag}), then again we are stuck with the troublesome relation
$f \gg M_P$. Therefore, we must demand that the total potential near the maximum is dominated 
by the higher order terms, e.g.,
\begin{equation}
\label{eq:quartic}
m^2 f^2 \ll |\lambda| f^4 \;;
\end{equation}
the potential is flattened near the maximum with respect to a conventional PNGB. 
On the other hand, a quadratically divergent contribution to the mass term
will be generated by the quartic coupling,
\begin{equation}
\label{eq:masscor}
\Delta m^2 = -\frac{3\lambda \, \Lambda^2}{16\pi^2} \; .
\end{equation}
Comparing it with (\ref{eq:quartic}), we see that the cutoff $\Lambda$ of
(\ref{eq:masscor}) must be much smaller than the naively expected
value $4\pi f$. In other words, there must be other fields with masses much
below $4\pi f$ which cut off the quadratic divergence. 
This is possible if their interactions soften the symmetry breaking due to the quartic term.
This requires some special structure of the theory 
which will be discussed in subsection~\ref{sec:little_higgs}.

Another possibility is that the total vacuum energy during inflation is
carried by another field as in hybrid inflation models~\cite{Linde:1993cn}.
The slow-roll field acts as a trigger of the phase transition of the other field.
In this case there is a similar worry that the coupling between these
fields, which is an explicit breaking contribution, can destroy the flatness of PNGB potential 
through radiative corrections. We will discuss
this case in subsection~\ref{sec:hybrid}.

\subsection{\label{sec:little_higgs}A model of single field 
inflation based on little Higgs theories}

From the discussion above we see that a single field inflation with
the field value smaller than $M_P$ requires that the quadratic term is smaller
than that naively induced by higher order terms. This is quite similar to what 
happens for the Higgs potential, so that we can use the same ideas of little
Higgs theories~\cite{Arkani-Hamed:2001nc,Arkani-Hamed:2002pa,Arkani-Hamed:2002qx,Arkani-Hamed:2002qy,Low:2002ws,Gregoire:2002ra}, recently proposed as a new solution to the hierarchy problem in 
electroweak symmetry breaking.

Let us first describe the general feature of the little Higgs
theories. For specific models we refer the reader to the
literature~\cite{Arkani-Hamed:2001nc,Arkani-Hamed:2002pa,Arkani-Hamed:2002qx,Arkani-Hamed:2002qy,Low:2002ws,Gregoire:2002ra}.
A little Higgs model is based on a chiral Lagrangian from some
spontaneously broken global symmetry. This symmetry is also
explicitly broken by two (or more) sets of couplings. Each set of
couplings preserves a different subset of the global symmetry under
which the little Higgs is an exact Nambu-Goldstone boson. The little
Higgs only learns its PNGB nature in the presence of both sets of
couplings when the symmetry is completely broken. Therefore, there is
no one-loop quadratically divergent contribution to the little Higgs
mass.  On the other hand, the quartic coupling can be generated at 
tree-level combining both sets of couplings.  The potential
for all little Higgs models has the following form
\begin{equation}
\label{quartic}
c_1 g_1^2 f^2 \left| \phi + i \frac{h^2}{f} \right|^2 + c_2 g_2^2 f^2
\left| \phi - i \frac{h^2}{f} \right|^2,
\end{equation}
where $f$ is the symmetry breaking scale, $g_1$ and $g_2$ represent
the two sets of couplings, $c_1,\, c_2$ are order 1 constants, $h$
is the little Higgs, and $\phi$ is a ``fat'' Higgs which receives large
contributions to its mass both from $g_1$  and $g_2$ alone.  
The first term preserves a shift symmetry
\begin{equation}
h \to h+\epsilon, \quad \phi \to \phi - \frac{2\, i\, \epsilon\, h}{f} \; ,
\end{equation}
while the second one preserves a different symmetry,
\begin{equation}
h \to h+\epsilon, \quad \phi \to \phi + \frac{2\, i\, \epsilon\, h}{f} \;;
\end{equation}
each one forbids a mass term for $h$.
For $c_1 g_1^2 +c_2 g_2^2>0$, we can integrate out the heavy 
$\phi$ field and obtain a quartic coupling for $h$,
\begin{equation}
\frac{4 c_1 c_2 g_1^2 g_2^2}{c_1 g_1^2+ c_2 g_2^2} |h|^4 \;.
\end{equation}
In addition, the radiative contribution to $h$ mass squared is
\begin{equation}
\frac{c_3 g_1^2 g_2^2}{16\pi^2}\, f^2 \;,
\end{equation}
with $c_3={\cal O}(1)$.  The coefficients $c_1, \, c_2,\, c_3$ can be
either positive or negative depending on the model and the types of
interactions.  In little Higgs theories of electroweak symmetry
breaking, one requires $c_1, \, c_2 >0$ and $c_3 <0$. To obtain a
model for inflation we make a different choice, $c_1 \cdot
c_2 <0$, $c_3<0$, so that both the squared mass term and the quartic
coupling are negative.  In terms of the inflaton $\psi$, which is
assumed to be the real part of $h$, $\psi = \sqrt{2} \Re (h)$, the
potential is
\begin{equation}
V(\psi) = V_0 -\frac{m^2}{2} \psi^2 - \frac{\lambda}{4} \psi^4 \;,
\end{equation}
where
\begin{equation}
m^2 = \frac{|c_3| g_1^2 g_2^2}{16\pi^2}\, f^2, \qquad \lambda=
\left|\frac{4 c_1 c_2 g_1^2 g_2^2}{c_1 g_1^2+ c_2 g_2^2}\right|, \qquad
V_0 \approx \lambda f^4 \;.
\end{equation}
Generically one expects a cutoff of the order of $4 \pi f$, therefore the quadratic term is 
suppressed by a loop factor with respect to that naively induced by the quartic coupling.
The form of the inflaton potential is similar to the one proposed in
Ref.~\cite{Dine:1997kf,Riotto:1997iv} in a gauge mediated SUSY
breaking model (though the $\eta$-problem was not addressed there).  Note that
the real potential is not unbounded from below because $\psi$ is a PNGB; the true 
minimum occurs at $\psi \sim f$.

The Universe is assumed to start near $\psi=0$. In the beginning, when
the $m^2\psi^2$ dominates the tilt of the potential, the Universe undergoes slow-roll inflation. 
Inflation ends after $\psi$ grows and the $\lambda\psi^4$ term becomes dominant. 
To be specific, we will assume that $c_1>0$, $c_2<0$, and $|c_2| g_2^2 \ll c_1 g_1^2$,
then $\lambda \approx 4|c_2| g_2^2$. During inflation the slow-roll
parameter $\eta$ is given by
\begin{equation}
\eta = \frac{M_P^2 V''}{V} \approx -\frac{g_1^2}{64\pi^2} \,
\left|\frac{c_3}{c_2}\right| \, \frac{M_P^2}{f^2} \;.
\end{equation}
The observational constraint $|\eta| < 1/20$ requires
\begin{equation}
\frac{f}{M_P} > \frac{g_1}{4\pi} \sqrt{\left|\frac{5c_3}{c_2}\right|} \;.
\end{equation}
It can be satisfied with $f < M_P$ if $g_1 \lesssim {\cal O}(1)$.
One can easily check that with this choice of parameters also the other 
slow-roll parameter $\epsilon$ is small.  
The number of e-foldings the Universe expands after the
$\lambda\psi^4$ term dominates is roughly given by $|\eta|^{-1}$, 
so if we assume that the COBE scale occurs when the $m^2\psi^2$ term is still more
important, $|\eta|$  can not be too small and should be close to the
current limit. In the opposite limit only the quartic term is relevant during observable
inflation. The tilt of the spectral index $n-1 \approx
2\eta$ in this model is predicted to be negative. From the COBE normalization for curvature
perturbation, we have
\begin{equation}
5.3\times 10^{-4} = \frac{V^{3/2}}{M_P^3 V'} \approx |\eta|^{-3/2}
\frac{m}{\psi} \approx |\eta|^{-3/2} \sqrt{\lambda} \;,
\end{equation}
where the last relation is obtained because the COBE scale should be near
the point where the $m^2 \psi^2$ term and the $\lambda \psi^4$ term
are comparable. This requires
\begin{equation}
g_2 \approx 2.7\times 10^{-3} |\eta|^{3/2} |c_2|^{-1/2} \;.
\end{equation}

The very small coupling $g_2$ can be seen as a weak point of this model,
though it is natural in the 't~Hooft's sense~\cite{'tHooft:1979bh}, 
because a larger symmetry is recovered in the limit $g_2 \to 0$. 
Another concern is whether the flat potential is preserved in the presence of quantum gravity effects.
One expects that, in addition to the couplings $g_1$ and $g_2$,
higher dimensional operators generated by quantum gravity effects may also break the global symmetry 
explicitly and give rise to a potential for the PNGB. 
These effects are suppressed by powers of $f/M_P$.
Which higher dimensional operators can be generated depends on the specific 
little Higgs theory and its UV completion as discussed in the previous section.

\subsection{\label{sec:hybrid}Hybrid inflation models}

In this subsection we consider 
hybrid inflation models with the inflaton being a PNGB\footnote{Two-field inflation with a  
PNGB triggering a first order phase transition has been proposed \cite{Adams:1990ds}, 
but the problems we are going to discuss are not addressed.}
In these models the slow-rolling  field is protected by an
approximate symmetry, while the vacuum energy is dominated by another
(waterfall) field. The first field acts like a trigger on
the other one: when a critical value is reached we are quickly  driven
to the true vacuum and inflation ends. 

At first sight, however, this
introduces another problem:  how can the approximate symmetry protect
the flatness of the potential without suppressing the coupling
between the slow-rolling field and the other one? Another way of
phrasing this problem is that since the coupling between the two fields
breaks the global symmetry, one may worry that it generates a large 
potential for the inflaton and spoil slow-roll inflation. 

Assuming that the slow-rolling inflaton $\psi$ and the waterfall field $\phi$
couple through the interaction
\begin{equation}
\lambda \psi^2 \phi^2 \;,
\end{equation}
this will generate a correction to the mass of the inflaton field,
\begin{equation}
\label{eq:masscor1}
\Delta m_{\psi}^2 \sim \frac{\lambda}{16\pi^2} \Lambda^2 \;,
\end{equation}
where $\Lambda$ is the cutoff of the integral. 
In order for $\psi$ to act as a switch on the waterfall field, we need
\begin{equation}
\lambda \psi_0^2 > |m_{\phi}^2| \;,
\end{equation}
where $\psi_0$ is the initial value of $\psi$.
This implies
\begin{equation}
\label{eq:masscor2}
m_{\psi}^2 > \frac{1}{16\pi^2}\, \frac{\Lambda^2 |m_{\phi}^2|}{\psi_0^2} \;.
\end{equation}
We can see that the cutoff $\Lambda$ can not be too high. 
The hybrid inflation requires $m_{\psi}^2 \ll |m_{\phi}^2|$.
It would not work with a naive
cutoff $\Lambda \sim 4\pi f$ expected in strong dynamics, which would yield
$m_{\psi}^2 > |m_{\phi}^2|$. 

Therefore, we need a much lower cutoff for the corrections to $m_{\psi}^2$.
The only known ways to have such a low cutoff are supersymmetry and 
little Higgs 
theories. 
In these cases, one may (at best) cut off the integral at $\Lambda^2
\sim |m_{\phi}^2|$, then we have
\begin{equation}
\label{eq:masscor3}
m_{\psi}^2 > \frac{1}{16\pi^2}\, \frac{m_{\phi}^4}{\psi_0^2} \;.
\end{equation}
On the other hand, the current constraint on the slow-roll parameter
$|\eta|<0.05$ implies
\begin{equation}
\label{eq:slow-roll_mass}
m_{\psi}^2 < 0.05 \times \frac{V}{M_P^2} \;.
\end{equation}
Comparing the above two equations, the requirement $\psi_0 \ll M_P$ implies $m_{\phi}^4 \ll V$;
the waterfall field $\phi$ has to be light compared to the scale of the total vacuum energy it controls. 
To get a natural small mass for $\phi$ again requires some symmetry reason. In contrast with the
case of the slow-roll field, SUSY can protect the lightness of the 
waterfall field because we only need $|m_\phi| < V^{1/4}$, not $|m_\phi|
\ll H$. Another possibility is that also the waterfall field is a PNGB, protected by a shift symmetry.

From these general arguments, we see that we are led to very specific structures for any natural hybrid 
4d models of inflation, if all field values are required to be smaller than $M_P$.
Either we need both SUSY and PNGBs, or we need a little Higgs structure
with all relevant fields being PNGBs. We present two examples to demonstrate it explicitly.

\subsubsection{\label{sec:susymodel}A SUSY model}

The idea of SUSY hybrid inflation with the inflaton as a PNGB was discussed in
Ref.~\cite{Cohn:2000hc,Stewart:2000pa}, in the context of non-Abelian
discrete symmetries.
To illustrate our point and to clarify the requirements, let us study
a very simplified model. Consider the superpotential
\begin{equation}
W= \lambda_0 S (\psi_1^2 +\psi_2^2 -f^2) + \frac{\lambda_1}{2} \psi_1
 \phi^2 + \lambda_2 X(\phi^2-v^2) \;,
\end{equation}
with
\begin{equation}
\lambda_1^2 f^2 > 2 \lambda_2^2 v^2 \;.
\end{equation}
The first term preserves a U(1) symmetry which is spontaneously
broken. We can parametrize the flat directions as follows,
\begin{equation}
\psi_1 +i\psi_2 \equiv \sqrt{2}Q =  (f+\sigma)\, e^{i\chi/f} \;, \quad
\psi_1 -i\psi_2 \equiv \sqrt{2}\overline{Q} =  (f-\sigma)\, e^{-i\chi/f}
\;,
\end{equation}
where $\chi$ is the Nambu-Goldstone boson of the broken U(1) symmetry,
and $\sigma$ is the other
flat modulus due to SUSY. When SUSY is broken, $\sigma$ receives a
potential and we assume that it is stabilized at $\sigma=0$.
We will only consider the field $\chi$, which plays the role of 
the inflaton. For convenience, in
the following we will also use $\psi_1$ and $\psi_2$ to simply
represent their values along this direction,
\begin{equation}
\psi_1 = f \cos \left(\frac{\chi}{f}\right) \; , \quad
\psi_2 = f \sin \left(\frac{\chi}{f}\right) \; .
\end{equation}

The U(1) symmetry is also explicitly broken by the coupling $\lambda_1$.
For the moment we assume that this is the only explicit breaking effect
and that the K\"{a}hler potential preserves the
U(1) symmetry up to corrections of order
$\lambda_{1}^2/(16\pi^2)$. 
A potential is generated for $\chi$ due to the $\lambda_1$ coupling.
We assume the initial condition of the early Universe to be $\chi \approx 0$; this
forces $\phi=0$ because this field receives a large mass from
$\psi_1$. SUSY is broken by $F_X$ and the vacuum energy density is
\begin{equation}
V_0 \approx |F_X|^2 = \lambda_2^2 v^4 \; .
\end{equation}
There are two kinds of contributions which lift the potential of $\chi$.
First, supergravity induces a soft SUSY breaking mass of
order $H$ for every scalar ($\psi_1,\, \psi_2,\, \phi$). However,
because $\chi$ is a PNGB, it only receives a potential due to
the presence of the explicit breaking $\lambda_1$. The corresponding
contribution is loop-suppressed,
\begin{equation}
\label{eq:sugra-tilt}
m_\chi^2\,({\rm SUGRA}) \sim \frac{\lambda_1^2}{16\pi^2}\, 3H^2 \;.
\end{equation}
One can see that there is no SUGRA $\eta$-problem if $\lambda_1 \lesssim 1$ 

In addition to the corrections due to supergravity,
there is a direct Yukawa mediated contribution through a $\phi$ loop,
arising from the splitting of the spectrum of the $\phi$ supermultiplet 
due to $F_X$.
The potential receives $\chi$ dependence at one loop,
\begin{equation}
\label{eq:yukawa-tilt}
V(\chi) \approx V_0 \left(1 + \frac{\lambda_2^2}{4\pi^2} \ln
\frac{\lambda_1 \psi_1}{\mu} \right)
=V_0 \left(1 + \frac{\lambda_2^2}{4\pi^2} \ln
\frac{\lambda_1 \cos(\chi/f)}{\mu/f} \right) \;.
\end{equation}
The derivatives are easy to calculate,
\begin{eqnarray}
V'(\chi) &=& - V_0 \frac{\lambda_2^2}{4\pi^2}
\frac{\sin(\chi/f)}{f \cos(\chi/f)}
= - V_0 \frac{\lambda_2^2}{4\pi^2}
\frac{\psi_2}{f\psi_1}\, , \\
V''(\chi) &=& - V_0 \frac{\lambda_2^2}{4\pi^2}
\frac{1}{f^2 \cos^2(\chi/f)}
= - V_0 \frac{\lambda_2^2}{4\pi^2}
\frac{1}{\psi_1^2} \; .
\label{eq:susymasscor}
\end{eqnarray}
We see that $\chi$ is rolling in the right direction ($0\to \pi
f/2$), and eq.~(\ref{eq:susymasscor}) agrees with eqs.~(\ref{eq:masscor2}),
(\ref{eq:masscor3}) with the cutoff $\Lambda^2 \sim |m_{\phi}^2| = \lambda_2^2
v^2$.
The slow roll parameter $\eta$ is now
\begin{equation}
\eta = M_P^2 \frac{V''}{V_0}=
-\frac{\lambda_2^2}{4\pi^2}\,\frac{M_P^2}{\psi_1^2} \;.
\end{equation}
We have
\begin{equation}
\lambda_2 = 2\pi \sqrt{|\eta|} \frac{\psi_1}{M_P} \;.
\end{equation}
Given the current constraint $|\eta|<0.05$, $\psi_1 \ll M_P$ requires 
$\lambda_2 \ll 1$, which is equivalent to say that $\phi$ has to be light. 
In the limit $\lambda_2 \to 0$, the potential becomes
flat because SUSY breaking vanishes. However, in this model
there is no enhanced symmetry in the Lagrangian in the $\lambda_2 \to 0$ limit.
This is technically natural in SUSY theories though because of the
non-renormalization theorem. For a non-SUSY theory the smallness of $\lambda_2$ would
be unstable against radiative corrections from the other interactions. 

Let us examine the other constraints. We will assume that the Yukawa mediated contribution,
eq.~(\ref{eq:yukawa-tilt}), dominates over the supergravity contributions,
eq.~(\ref{eq:sugra-tilt}). The number of e-folds of slow-roll inflation after a given epoch is
\begin{equation}
N(\chi) \simeq \int_{\chi_{\rm end}}^{\chi} M_P^{-2} \frac{V}{V'}d\chi
= \int_{\psi_{1{\rm end}}}^{\psi_1(\chi)} M_P^{-2}
\frac{d\psi_1}{\frac{\lambda_2^2}{4\pi^2}\frac{1}{\psi_1}}
= \frac{4\pi^2}{\lambda_2^2 M_P^2} \left( \frac{\psi_1^2}{2} -
\frac{\psi_{1{\rm end}}^2}{2} \right)
= \frac{1}{2|\eta(\chi)|}- \frac{1}{2|\eta_{\rm end}|} \;,
\end{equation}
where the subscript ``end'' represents the end of inflation.
Generally it is dominated by the first term. We obtain a prediction
for the deviation of the spectral index from 1
\begin{equation}
n-1 \simeq 2\,\eta_{\rm COBE} \approx -\frac{1}{N_{\rm COBE}} \;,
\end{equation}
where the subscript ``COBE'' denotes the values corresponding to the scale of COBE measurement,
and $N_{\rm COBE}$ is typically $40-60$.

From the constraint on curvature perturbation measured by COBE,
\begin{equation}
5.3\times 10^{-4}= \frac{V^{3/2}}{M_P^3 |V'|} \approx
\frac{1}{M_P^3}\,\frac{\lambda_2^3 v^6}{
\lambda_2^2 v^4 \frac{\lambda_2^2}{4\pi^2}
\frac{\psi_2}{f\psi_1}}
= \frac{2\pi v^2 f}{M_P^2 \psi_{2}} \, |\eta|^{-1/2}
= \frac{2\pi v^2 }{M_P^2 \sin( \chi /f)} \, |\eta|^{-1/2} \;.
\end{equation}
we obtain
\begin{equation}
\frac{v^2}{M_P^2} \approx 8\times 10^{-6} \sqrt{\frac{50}{N_{\rm COBE}}}
\sin \left( \frac{\chi_{\rm COBE}}{f}\right) \quad
\Rightarrow \frac{v}{M_P} \approx 3\times 10^{-3}
\left(\frac{50}{N_{\rm COBE}}\right)^{\frac{1}{4}}
\sin^{\frac{1}{2}} \left( \frac{\chi_{\rm COBE}}{f}\right) \;.
\end{equation}

In this simple model we did not address why there is an approximate
U(1) global symmetry, broken only by $\lambda_1$. We left out many other possible terms which
do not respect the U(1) symmetry. While it is technically natural
in SUSY theories, it is not very well motivated.
In addition, explicit breaking terms can also arise from quantum gravity effects.
Our main point here is to demonstrate how
an approximate shift-symmetry can protect a flat direction from
SUGRA corrections during inflation and the subtleties involved
when one needs to couple the PNGB to other fields for the end of
inflation.
One can extend the model in such a way that a non-Abelian discrete symmetry
gives rise to the approximate global symmetry as
in~\cite{Cohn:2000hc,Stewart:2000pa}, and suppresses any dangerous symmetry breaking terms.
Another natural way to obtain an accidental global symmetry is to exploit the locality in
(deconstructed) extra dimensions, which we will study next. In particular,
one can find natural hybrid models even without SUSY.

\subsubsection{\label{sec:6d}A 6d hybrid model and its 4d deconstruction}

In this subsection we present a non-SUSY hybrid inflation model.
As we argued earlier, to get such a model in 4d without SUSY, 
we need the little Higgs structure with both
the inflaton field and the waterfall field being PNGBs.
Little Higgs theories were first motivated from deconstructing
extra-dimensional theories, 
where the PNGBs correspond to the extra components of gauge fields
in extra dimensions~\cite{Arkani-Hamed:2001nc,Arkani-Hamed:2002pa}.
We will first discuss a 6d model where the extra components of the 
gauge field, $A_5$, $A_6$, play the roles of the inflaton and waterfall
fields. It provides simple physics intuition and a clear picture 
of inflation dynamics. Later we will show that it can safely be 
deconstructed to purely 4d models.

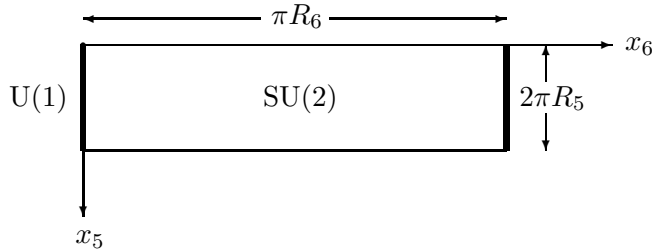
\begin{figure}
\caption{\label{fig:6D} The 6d orbifold model.}
\begin{picture}(300,120)(-50,-80)
\put(0,0){\vector(1,0){200}} \put(205,-2){$x_6$}
\put(0,0){\vector(0,-1){65}} \put(-3,-75){$x_5$}
\put(0,0){\circle*{2}} \put(0,-40){\line(1,0){160}}
\put(68,-23){SU(2)}
\put(-28,-23){U(1)} \put(71,8){$\pi R_6$}
\put(67,10){\vector(-1,0){67}} \put(93,10){\vector(1,0){67}}
\put(165,-23){$2\pi R_5$} \put(175,-14){\vector(0,1){14}}
\put(175,-26){\vector(0,-1){14}} \linethickness{2pt}
\put(0,0){\line(0,-1){40}}
\put(160,0){\line(0,-1){40}}
\end{picture}
\end{figure}

We consider an SU(2) gauge theory in 4 ordinary infinite dimensions
and 2 extra compact dimensions. We assume that the $x_5$ direction is
compactified on a circle with radius $R_5$, and the $x_6$ direction is
compactified on an $S^1/Z_2$ orbifold with radius $R_6$ (Fig.~\ref{fig:6D}). 
Furthermore we assume that the orbifold projection breaks the SU(2) gauge 
symmetry down to U(1) at the orbifold fixed points $x_6 = 0, \pi R_6$.  
The assumed parities of the various gauge components under the $Z_2$ projection 
are shown in Table~\ref{tab:parities}.
\begin{table}
\caption{\label{tab:parities} The $Z_2$ parities of various gauge
components.}
\begin{tabular}{c|ccc}
 & $A_{\mu}^a$ & $A_5^a$ & $A_6^a$ \\ \hline $T^{1,2}$ & $-$ & $-$ & +
\\ $T^3$  & + & + & $-$
\end{tabular}
\end{table}
We see that in the 4d picture, the $T^3$ component of $A_5$ and
$T^{1,2}$ components of $A_6$ have zero modes. They have a tree level
potential from the commutator term in $F_{56}^2$,
\begin{equation}
\label{commutator}
\frac{1}{2}g^2 (A_5^3)^2 \left[(A_6^1)^2+ (A_6^2)^2\right] \;,
\end{equation}
where $g$ is the 4d gauge coupling.
For simplicity, we used $A_5$, $A_6$ to represent the zero modes
and we will omit the generator indices in the rest of the discussion.  Note
that the full theory is periodic under the transformations $A_5 \to
A_5+1/(gR_5)$, $A_6 \to A_6+ 1/(gR_6)$, where the Kaluza-Klein modes
simply shift by one unit.

At tree level, there are flat directions along $A_5$ (with $A_6=0$),
and $A_6$ (with $A_5=0$). 
As discussed in section~\ref{sec:idea}, $A_{5,6}$ can not have
local mass terms by gauge invariance. 
They can only get non-local contributions from Wilson
lines and these contributions
are indeed generated by radiative corrections, lifting the flat directions.
One can imagine that at the origin
($A_5=A_6=0$) the  radiative corrections generate a positive squared
mass for one direction, say $A_6$, and a (larger) negative mass squared for the other
($A_5$).  This can be achieved if there are charged fermions living at
the orbifold fixed lines, $x_6=0$ or $\pi R_6$.
In this case $A_6$ can play the role of the
slow-roll field and $A_5$ can be the waterfall field. The Universe
with the initial condition $A_6 \neq 0,\, A_5=0$ will slowly roll to
the origin until the squared mass of $A_5$ turns negative and $A_5$
jumps down to the true vacuum. To satisfy the slow-roll condition
during inflation and to have a sufficiently fast waterfall process it
is required that
\begin{equation}
\label{m5m6}
m_{A_6}^2 \ll H^2 \ll \left|m_{A_5}^2\right| \;,
\end{equation}
which implies $R_6 \gg R_5$.

Similarly to the case of one extra dimension in sec.~\ref{sec:idea},
the potential can be computed for a given particle content, and is
well known in the literature. A more detailed discussion of
the parameters required for slow-roll inflation and all 
the other observational constraints can be found in the Appendix.
The most important feature of this model is again that the locality in
extra dimensions protects the flat potential for the inflaton
(and also the mass of the waterfall field) while still allowing significant
couplings which trigger the waterfall phase transition and reheat the
Universe after inflation.

From the constraints on the parameters discussed in the Appendix,
one can check that the effective decay constants or symmetry breaking
scales $1/(2\pi g R_5)$, $1/(2\pi g R_6)$ can be smaller than $M_P$
in this model. Therefore, a valid 4d model can easily be obtained 
by deconstruction. In fact, there is more freedom
in the 4d deconstructed theories, since the various couplings are not
required to be related as in the 6d theory by the higher dimensional 
gauge symmetry.
The hierarchy between the scales of the inflaton field and the waterfall
field can either come from the symmetry breaking
scales or the couplings. 

In the 4d picture, the inflaton and the waterfall fields are PNGBs,
whose masses are protected by many approximate symmetries.
Because in the limit in which one
of them is restored the PNGB is exactly massless, its mass can be
quite small even in the presence of a large coupling to additional fields. 
\begin{figure}\begin{center}
\begin{picture}(250,66)
\BCirc(47,33){17} \BCirc(207,33){17} \ArrowArcn(127,-53)(120,125,55)
\ArrowArcn(127,-123)(173,111.8,68.2) \ArrowArc(127,119)(120,235,305)
\ArrowArc(127,189)(173,248.2,291.8) \Text(137,75)[]{\large X}
\Text(143,71)[]{1} \Text(137,42)[]{\large X} \Text(143,38)[]{2}
\Text(113,26)[]{\large X} \Text(119,22)[]{3} \Text(113,-7)[]{\large X}
\Text(119,-11)[]{4} \Text(47,33)[]{SU(2)} \Text(208,33)[]{U(1)}
\end{picture} \vspace{0.5cm}
\caption{\label{fig:decon}\small The moose diagram for the
deconstructed model of hybrid inflation.}
\end{center} \end{figure}
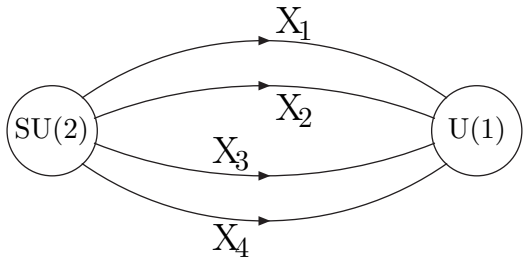
Let us start from two sites with an SU(2) symmetry on each site,
with four link fields $X_i$, $i=1,2,3,4$ (see fig.~\ref{fig:decon}), which transform as 
fundamentals under both SU(2)'s: the VEVs of the $X_i$ break SU(2) $\times$ SU(2) to the
diagonal  subgroup. 
To reproduced the orbifold projection we gauge the full SU(2) group on the first
site but only the U(1) subgroup corresponding to the $T^3$ generator
on the second site.
We add to the potential two plaquette operators:
\begin{equation}
\label{eq:plaquette}
V = - \kappa_1 f^4 {\rm Tr}(X_1 X_2^\dagger X_3 X_4^\dagger) -
\kappa_2 f^4 {\rm Tr}(X_2 X_3^\dagger X_4 X_1^\dagger) \;.  \end{equation}
Gauge fixing $X_1 = 1\!\! 1$, it is easy to find that the classically
flat directions can be  parameterized by $X_2$ and $X_4$, ($X_3 = X_2
X_4$), with the additional constraint $X_2 X_4 = X_4 X_2$,  coming
from the second plaquette
\cite{Arkani-Hamed:2002qx,Gregoire:2002ra}. In this way we have
reproduced  the commutator potential between $A_5$ and $A_6$. 
As only $T^3$ is gauged in the second site, we can add plaquette operators
which include the projection $\Omega \equiv {\rm diag}(1,-1)$ on this site
to get rid of useless light states. The operator
\begin{equation}
\label{eq:omega4}
- \kappa_3 f^4 {\rm Tr}(X_1 \Omega X_2^\dagger X_3 \Omega
X_4^\dagger) \;, 
\end{equation}   
forces $X_4$ to commute with $\Omega$ so that only the neutral component 
survives, and
\begin{equation}
\label{eq:omega2}
- \kappa_4 f^4 {\rm Tr}(X_1 \Omega X_3^\dagger X_4 \Omega
X_2^\dagger) \;, 
\end{equation}  
keeps only the charged components of
$X_2$. 
It is easy to verify that there are no one-loop quadratically divergent
contributions which lift the flat directions
\cite{Arkani-Hamed:2002qx,Gregoire:2002ra}:  each of the plaquette
interactions and gauge couplings respect a subgroup of the ${\rm
SU(2)}^8$ global  symmetry of the link fields, leaving the PNGBs
exactly massless. Only the combination of  two sources of explicit
breaking lift the flat directions, so that we have only logarithmic
divergences.

As in the 6d model the charged scalars from $X_2$ take a positive mass
squared, from gauge loops: $m_2^2 \sim g^4/(16 \pi^2) f^2$.
Instead we want  that the neutral component of $X_4$ receives
a tachyonic contribution to the potential: this can be  done
by introducing fermions in the theory, coupled to $X_4$. To avoid the
presence of quadratic divergences,  fermions must be introduced in a
``delocalized way'', similarly to what one does to control the top
loop  corrections to the Higgs mass in little Higgs models. We get a
contribution:  $m_4^2 \sim - \lambda^4 /(16 \pi^2) f^2$,
where $\lambda$ is the analogue of the top  Yukawa
coupling.
The required hierarchy can be obtained if $\lambda \gg g$.
Inflation can evolve as in the 6d model: we start away from the
origin along the $X_2$ direction and we roll towards the origin,
because of the positive mass contribution. For sufficiently big values
of $X_2$, $X_4$ is stuck at the origin because of the commutator
potential. When a critical point is reached, the potential of $X_4$
becomes unstable and the waterfall process starts, ending inflation.

The model is quite similar to the 6d one, except that there are 
differences due to couplings and volume factors, and to the fact that now gravity is 4-dimensional.  
As gravity is now 4-dimensional, one needs to worry about the corrections of the potential
due to explicit symmetry breaking operators generated by quantum gravity effects (which were exponentially
suppressed in ``real'' extra dimensions). For the simplest 2-site model, there are gauge invariant 
operators involving only two links. If one imagines that the links come from bilinear fermion condensates, 
they correspond to dimension-6 operators which are still dangerous as discussed before. These operators 
can be eliminated by additional (continuous or discrete) gauge symmetries\footnote{For example, if in the UV 
theory $X_1,\,X_2,\, X_3,\, X_4$ carry additional gauge charges 1, 3, 7, 5, respectively, the lowest dimensional 
gauge invariant operators which break the global symmetry are the plaquette operators in (\ref{eq:plaquette}). 
Note that ${\rm Tr} X_i X_i^\dagger$ does not break the global symmetry.}.
A deconstruction with more sites (gauge groups)
so that the gauge invariant operators require more links is also sufficient to
suppress quantum gravity effects~\cite{Hill:2002kq}.
As there is more freedom in the 4d deconstructed model, changing
the model parameters we can have a very fast waterfall process along
$X_4$ as in the original hybrid inflation \cite{Linde:1993cn} or a
quite slow one as in the SUSY inspired ``supernatural models''
\cite{Randall:1995dj}.

In the above discussion we assumed that $A_6$ ($X_2$) is the slow-roll field
and $A_5$ ($X_4$) is the waterfall field. One can also consider the opposite
case where the roles of the two fields are reversed. In that case one
requires $m_{A_5}^2 > 0$, $m_{A_6}^2 <0$, and $R_5\gg R_6$. Cosmic strings will 
be generated at the end of inflation because U(1) is broken; this process can have interesting 
but complicated consequences~\cite{Garcia-Bellido:1996qt}.

\section{\label{sec:conclusions} Conclusions} 

From the point of view of cosmology, inflation is definitely the most
attractive scenario describing the very early Universe. On the other hand,
from the particle physics point of view, the required inflaton
potential is extremely unnatural and seems to need a lot of fine tuning.
In this paper we examined in detail the physics ideas which may
be used to naturally obtain a viable inflaton potential. We emphasize that
SUSY, although being a popular paradigm for inflation models, can not
adequately preserve the flat potential for inflaton by itself.
The only natural way to obtain a flat potential for inflation is 
to incorporate some approximate shift symmetry. The examples are PNGBs
and extra components of gauge fields living in extra dimensions.
In purely 4d theories, the simplest model based on PNGBs has the difficulties
that we need to extrapolate the field theory beyond its regime of validity
as the symmetry breaking scale has to be greater than the Planck scale.
To avoid this problem we need either extra dimensions or more complicated
models in pure 4 dimensions as we discussed in sections~\ref{sec:idea}
and~\ref{sec:models}. 

With extra dimensions, locality in the extra space allows to get a trustworthy potential
even if the variation of the inflaton field is bigger than $M_P$, with exponentially
suppressed quantum gravity corrections \cite{Arkani-Hamed:2003wu}. 

On the other hand, purely 4d models require more sophisticated structures.
The inflaton, besides being a PNGB, must have a potential with further
protection from the potentially dangerous explicit symmetry breaking interactions 
which are required to end inflation. We conclude that supersymmetry
or little Higgs structure is necessary for such protection.
Discrete or continuous gauge symmetries are required in the UV completion
of the 4d models below the Planck scale to control quantum gravity effects.

A generic prediction of the 4d models is that the contribution
to the density perturbations from gravitational waves is unobservably
small, because the field values are smaller than $M_P$. This has to be
contrasted with extra-dimensional setups, where a significant production of
gravitational waves is possible.

In hybrid models density perturbations produced during the final phase transition 
can give interesting phenomenological signatures \cite{Garcia-Bellido:1996qt}.
 
The prediction of the spectral index depend on the individual models.
However, as it is quite difficult to preserve a flat direction during 
inflation, it seems quite generic that the spectral index $n$ should 
deviate from unity considerably. In our models the slow-roll parameters
are small because of loop-factor suppressions, so that we do not expect 
them to be {\em utterly} small.
Anyway, the same conclusion holds if the $\eta$-problem is solved just by a certain amount of fine-tuning.
Therefore, a small deviation from scale-invariance seems to be a smoking gun for the inflationary paradigm itself. 
In some sense the inflaton mass cannot be too separated from the Hubble scale during inflation 
for the same reason we do not expect the Higgs mass to be very far from the scale of new physics (whatever it is), 
in which the SM is embedded.

\vspace{1cm} {\bf Note added}: As this work was completed ref.~\cite{DN} appeared, where hybrid
models based on PNGBs are discussed.

\appendix

\section{More details about the 6 dimensional hybrid model and its deconstruction}

In this Appendix we present more detailed discussion of the 6d hybrid model
and its 4d deconstruction in sec.~\ref{sec:6d}.

For the 6d model,
the potential for the extra components of the gauge field  
consists of a sum of cosine functions which are periodic in $A_5 \to
A_5+1/(gR_5)$ and  $A_6 \to A_6+ 1/(gR_6)$. Here we will just expand
it around the origin to obtain the mass terms for $A_5$ and $A_6$ at
the origin.  To satisfy eq.~(\ref{m5m6}) we have to assume that $R_5
\ll R_6$.  The gauge loops give positive contributions to the squared
mass for both $A_5$ and $A_6$. For $R_5\ll R_6$ they are given by
\begin{eqnarray}
m_{A_5}^2 ({\rm gauge}) &\approx& \frac{2 g^2 \zeta(4)}{\pi^5}\,
\frac{R_6}{R_5^3}, \\ m_{A_6}^2 ({\rm gauge}) &\approx& \frac{3 g^2
\zeta(3)}{4\pi^4}\, \frac{1}{R_6^2} \;.
\end{eqnarray}
To make $m_{A_5}^2$ negative, we can introduce fermions extended along
the  $x_5$ direction but localized in the $x_6$ direction, so that
they only contribute to the mass of $A_5$. A natural choice is to have
charged fermions living at the orbifold fixed lines, $x_6=0$ or $\pi R_6$, 
which preserves only the  U(1) gauge symmetry (corresponding to
$A^3$). Their contribution  to $m_{A_5}^2$ is\footnote{In fact,
this contribution is localized at the orbifold fixed lines, which
can mix KK modes in the $x_6$ direction. However, as we see later, this
term is required to be smaller than the tree-level KK masses $n^2/R_6^2$,
so we can treat it as a small perturbation without re-diagonalizing 
the mass eigenstates.}
\begin{equation}
m_{A_5}^2 ({\rm fermion}) \approx -\frac{3 g^2 \zeta(3)}{4\pi^4}\,
\frac{1}{R_5^2} \sum_i 2 Q_i^2 \;,
\end{equation}
where $Q_i$ is the $U(1)$ charge of the fermion $i$.  From the
constraint (which will be discussed later) on the large  density
perturbations generated at the beginning of waterfall, $R_6/R_5$ is
required to be  $\gtrsim 20$. We can see that for  somewhat large
$Q_i$ or many fermions, $m_{A_5}^2$ can become negative. The
exact values of $m_{A_5}^2,\, m_{A_6}^2$ depends on the field
content. Nevertheless, they have to be cut off by $1/R_5^2$ and $1/R_6^2$
as they should vanish in the $R_{5,6} \to \infty$ limit.
Below we will simply parametrize them by
\begin{eqnarray}
m_{A_5}^2  &=& -\frac{c_5 g^2 }{\pi^4}\, \frac{1}{R_5^2}, \\ m_{A_6}^2
&=& \frac{c_6 g^2 }{\pi^4}\, \frac{1}{R_6^2} \;,
\label{eq:ma6}
\end{eqnarray}
where $c_5$ and $c_6$ are constants of order 1 or larger.
Note that we require that the positive squared mass for  $A_5$ from
the tree-level potential eq.~(\ref{commutator}) to be larger than the
negative radiative contribution in the beginning of
inflation, i.e.,
\begin{equation}
\label{eq:initialmass}
|m_{A_5}^2| = \frac{c_5 g^2}{\pi^4 R_5^2} < g^2 A_6^2 \,({\rm
initial}) \lesssim \frac{1}{4R_6^2} \quad \Rightarrow \quad R_5 >
\frac{2\,\sqrt{c_5}\,g}{\pi^2}\, R_6 \; .
\end{equation}
We can see that 
eqs.(\ref{eq:ma6}), (\ref{eq:initialmass}) are consistent 
with eqs.(\ref{eq:masscor2}), (\ref{eq:masscor3}) as 
$\Lambda^2 \sim 1/R_6^2 > |m_{A_5}^2|$.

During inflation, the vacuum energy is dominated by the $A_5$
potential,
\begin{equation}
V \sim \frac{|m_{A_5}^2|}{g^2 (2\pi R_5)^2} \sim \frac{c_5}{4\pi^6}\,
\frac{1}{R_5^4} \;,
\end{equation}
while the slope is determined by the $A_6$ potential,
\begin{equation}
V' \sim \frac{m_{A_6}^2}{g(2\pi R_6)} \sim \frac{c_6 g}{2\pi^5}\,
\frac{1}{R_6^3}, \quad \quad  V'' \sim m_{A_6}^2 \;.
\end{equation} 
There are several constraints for the parameters $g$, $R_5$, $R_6$.
For the slow roll condition, we require
\begin{equation}
\eta = M_P^2 \frac{V''}{V} \approx g^2 (2\pi R_5)^2 M_P^2 \,
\frac{c_6}{c_5} \, \frac{R_5^2}{R_6^2} \ll 1 \;.
\end{equation}
From the COBE measurement of curvature perturbations, it is required
that
\begin{equation}
5.3\times 10^{-4} = \frac{V^{3/2}}{M_P^3 V'} \approx \eta^{-3/2}
c_6^{1/2} \frac{2}{\pi}\, g^2 \;.
\end{equation}

Finally, as discussed in Ref.~\cite{Randall:1995dj}, in hybrid
inflation models, large density perturbations will be generated during
the period when both fields are light. A rough condition for a fast enough 
end of inflation, so that the large density perturbations do not fall inside 
the observable window is
\begin{equation}
\frac{m_{A_5}}{H}\, \frac{m_{A_6}}{H} \approx \eta
\sqrt{\frac{c_5}{c_6}} \,\frac{R_6}{R_5} \gtrsim 1 \;.
\end{equation}

For some sample numbers, if we assume $c_5,\, c_6 \sim 1$,  all
constraints can be satisfied with
\begin{equation}
\frac{R_6}{R_5} \sim 100, \quad g \sim 2\times 10^{-3},\quad 2 \pi R_5
M_P \sim 10^4, \quad \eta \sim 0.03 \;.
\end{equation}
One can check that the effective decay constants or symmetry breaking
scales $1/(2\pi g R_5)$, $1/(2\pi g R_6)$ can be smaller than $M_P$
in this model. Therefore, in this case it is possible to obtain 
valid 4d models by dimensionally deconstructing this model.

For 4d models there is more freedom in the choice of parameters
as couplings are not related by higher dimensional gauge symmetry.
For simplicity of discussion, we assume
$\kappa \sim g^2$ for the couplings of the 
plaquette operators (\ref{eq:plaquette}), where $g$ is the
SU(2) gauge coupling: in this way the commutator potential is $\sim
g^2$ as happens in the 6d model.
We also assume that the VEVs 
for $X_i$'s are equal ($\sim f$),
and the required hierarchy between the scales of the slow-roll
field and the waterfall field is generated by the couplings to fermions.

The charged scalars from $X_2$ have a positive mass
squared from gauge loops: $m_2^2 \sim g^4/(16 \pi^2) f^2$
(\footnote{With the assumption $\kappa \sim g^2$,  the combined
contribution of the plaquette operators is comparable to the gauge
term.}). The neutral scalar of $X_4$, on the other hand, receives
a negative contribution to its mass squared from couplings to fermions,
$m_4^2 \sim - \lambda^4 /(16 \pi^2) f^2$,
where $\lambda$ is the analogue of the top  Yukawa
coupling\footnote{These 1-loop contributions are slightly enhanced
with respect to the explicit UV operator, which can be estimated from
the two loop quadratic divergence through naive dimensional
analysis, by the logarithmic factor  $\log(\Lambda^2/(g f)^2) \simeq 2
\log(4\pi/g)$. We will see in the following that a viable model of
inflation requires a small $g$ so that this enhancement is quite
consistent, while it is rather small in models of EWSB, where $g \sim
1$.}. As emphasized, the fermion couplings should be introduced in a
``delocalized way,'' i.e., preserving enough global symmetries, to
avoid the quadratic divergences.
The required hierarchy can be obtained if $\lambda \gg g$.

Inflation evolves as in the 6d model: we start away from the
origin along the $X_2$ direction and we roll towards the origin. 
For sufficiently big values
of $X_2$, $X_4$ is stuck at the origin because of the commutator
potential. For this to happen, the positive contribution from the commutator
potential must be able
to overcome the negative one from the fermion couplings: 
\begin{equation}
\label{eq:ineqlambda}
g^2 f^2 > \frac{\lambda^4}{16 \pi^2} f^2 \;.  
\end{equation} 
The vacuum energy during the slow-roll is approximately constant and given by
$V_0 \sim \frac{\lambda^4}{16 \pi^2} f^4$.  Before we get to
the origin the mass along $X_4$ becomes negative and we end up to the
true minimum with restored gauge symmetry.  We can easily estimate the
slow-roll parameters: 
\begin{equation}
\label{eq:epsilondec}
\epsilon \equiv \frac{M_P^2}{2} \left(\frac{V'}{V_0}\right)^2 \simeq
\frac{M_P^2}{f^2}  \left(\frac{g^4}{\lambda^4}\right)^2 \;,
\end{equation} 
\begin{equation}
\label{eq:etadec}
\eta \equiv M_P^2 \frac{V''}{V_0} \simeq \frac{M_P^2}{f^2}
\frac{g^4}{\lambda^4} \;.  
\end{equation}
To have slow-roll we must assume that  
\begin{equation}
\label{eq:etadisin}
f \gg M_P \frac{g^2}{\lambda^2} \;.  
\end{equation} 
Note that,
assuming a certain hierarchy between the coupling constant $g \ll
\lambda$,  we have slow roll, even if $f$ is smaller than
the Planck scale, in constrast to what happens in  natural
inflation. This is possibile in a hybrid model because the vacuum
energy depends on  $\lambda$, while the slow-roll potential
is lifted by $g$.  Let us now look at the COBE normalization for the
large scale perturbations CMBR 
\begin{equation}
\label{eq:COBEdec}
\left(\frac{V_0}{\epsilon}\right)^{1/4} \simeq 0.027 \cdot M_P \;; 
\end{equation}
it gives the constraint 
\begin{equation}
\label{eq:COBEdeccon}
\frac{\lambda^3}{\sqrt{4 \pi} \cdot g^2}
\left(\frac{f}{M_P}\right)^{3/2} =  \frac{g \cdot
\eta^{-3/4}}{\sqrt{4\pi}} \simeq 0.027 \;.  
\end{equation} 
Using the experimental limit $\eta < 1/20 $, we get a rather strong constraint 
on $g$: $g \lesssim 0.01$.


\end{document}